\documentclass[12pt,preprint,nofootinbib]{revtex4}
\usepackage{times}
\usepackage{amsmath}
\usepackage{graphicx}
\usepackage{amssymb}

\begin{document}
%%%%%%%%%%
\def\D0{D\O~}
%%%%%%%%%%

\begin{flushright}
MSUHEP-040106\\
hep-ph/0401026
\end{flushright}
\title{Combined Effect of QCD Resummation and QED Radiative
Correction\\
to $W$ boson Observables at the Tevatron}

\author{Qing-Hong Cao} \email{cao@pa.msu.edu}
\author{C.--P. Yuan}\email{yuan@pa.msu.edu}

\affiliation{
\vspace*{2mm}
{Department of Physics \& Astronomy, \\
   Michigan State University, \\
     East Lansing, MI 48824, USA. \\ }}

\vspace{0.3in}

\begin{abstract}
A precise determination of the $W$ boson mass at the Fermilab Tevatron
requires a theoretical calculation in which the effects of the initial-state
multiple soft-gluon emission and the final-state photonic correction
are simultaneously included . Here, we present such a calculation
and discuss its prediction on the transverse mass distribution of
the $W$ boson and the transverse momentum distribution of its 
decay charged lepton, which are the most relevant observables for 
measuring the $W$ boson mass at hadron colliders. 
\end{abstract}
%12.20.-m Quantum electrodynamics
%12.38.-t Quantum chromodynamics
%2.38.Cy  Summation of perturbation theory
%12.15.Lk Electroweak radiative corrections
\maketitle

As a fundamental parameter of the Standard Model (SM), the mass of the
$W$-boson ($M_{W}$) is of particular importance. Aside from being
an important test of the SM itself, a precision measurement of $M_{W}$,
together with an improved measurement of top quark mass ($M_{t}$),
provides severe indirect bounds on the mass of Higgs boson
($M_{H}$). With a precision of 27 MeV for $M_{W}$ and 2.7 GeV for
$M_{t}$, which are the target values for Run II of the Fermilab Tevatron
collider, $M_{H}$ in the SM 
can be predicted with an uncertainty of about 35\%~\cite{Baur:2001yp}.
Comparison of these indirect constraints on $M_{H}$ with the results
from direct Higgs boson searches, at the 
LEP2, the Tevatron and the CERN
Large Hadron Collider (LHC), will be an important test of the SM. In order
to have a precision measurement of $M_{W}$, the theoretical uncertainties,
dominantly coming from the transverse momentum of the $W$-boson ($P_{T}^{W}$),
the uncertainty in parton distribution function (PDF) and the 
electroweak (EW) radiative
corrections to the $W$ boson decay, must be controlled
to a better accuracy~\cite{Baur:2000bi}.

At the Tevatron, about ninety percent of the production cross section
of $W$ boson is in the small transverse momentum region, where $P_{T}^{W}\leq20$
GeV. 
When $P_{T}^{W}$ is much smaller than $M_{W}$,
every soft-gluon emission will induce a large
logarithmic contribution to the $P_{T}^{W}$ distribution so that the
order-by-order perturbative calculation
in the theory of Quantum chromodynamics (QCD) 
cannot accurately describe the
$P_{T}^{W}$ spectrum 
and the contribution from multiple soft-gluon emission,
which contributes to all orders in the expansion of the strong coupling
constant $\alpha_s$, needs to be summed to all orders.
It has been shown that by applying the renormalization group
analysis, the multiple soft-gluon radiation effects can be resummed
to all orders to predict the $P_{T}^{W}$ distribution that agrees 
with experimental data~\cite{Balazs:1995nz, Balazs:1997xd}. RESBOS,
a Monte Carlo (MC) program ~\cite{Balazs:1997xd} 
resumming the initial-state soft-gluon radiations
of the hadronically produced lepton pairs through EW vector boson
production and decay at hadron colliders 
$p\bar{p}/pp\rightarrow V(\rightarrow\ell_{1}\bar{\ell_{2}})X$,
has been used by the CDF and \D0 Collaborations at the Tevatron to
compare with their data in order to determine $M_W$.
However, RESBOS does not include any 
higher order EW corrections to describe the vector boson decay. The EW
radiative correction, in particular the final-state QED 
correction, is crucial for precision measurement of $W$ boson mass
at the Tevatron, because photon emission from the final-state charged
lepton can significantly modify the lepton momentum which is used in
the determination of $M_{W}$. In the CDF Run Ib $W$ mass measurement,
the mass shifts due to radiative effects were estimated to be $-65\pm20$
MeV and $-168\pm10$ MeV for the electron and muon channels, respectively~\cite{Affolder:2000bp}.
The full next-to-leading order (NLO) $O(\alpha)$ EW 
corrections have been calculated~\cite{Dittmaier:2001ay, Baur:1998kt}
and resulted in WGRAD~\cite{Baur:1998kt}, a MC program
for calculating $O(\alpha)$ EW radiative corrections to the process
$p\bar{p}\rightarrow\nu_{\ell}\ell(\gamma)$. However, WGRAD does not include
the dominant correction originated from the initial-state multiple soft-gluon
emission. To incorporate both the initial-state QCD and and final-state
QED corrections into a parton level MC program is urgently required
to reduce the theoretical uncertainties in interpreting the experimental
data at the Tevatron. It was shown in Refs. \cite{Dittmaier:2001ay, Baur:1998kt} that at the NLO,
the EW radiative correction in $p\bar{p}\rightarrow\ell\nu_{l}(\gamma)$ is
dominated by the final-state QED (FQED) correction. Hence, in this
paper we present a consistent calculation which includes both the
initial-state multiple soft-gluon QCD resummation and the final-state
NLO QED corrections, and develop an upgraded version of 
the RESBOS program,
called RESBOS-A~\cite{RESBOS-A}, to simulate the signal events. 

The fully differential cross section for the production and decay
of the $W$ boson that includes only the effect of the 
initial-state multiple QCD soft-gluon emission
can be found in Ref.~\cite{Balazs:1997xd}.
To include also the final-state NLO QED contributions, we sum up 
the following two sets of differential cross sections. One, cf.
Eq.~(\ref{eq:ResFor}), contains final-state QED virtual correction
and part of the real photon emission contribution in which photon is either
soft or collinear. Another, cf. Eq.~(\ref{eq:ResFor2}), includes 
the hard photon contribution from the 
real photon emission processes.
Denote $Q$, $y$, $Q_T$ and $\phi_W$ to be the invariant mass, 
rapidity, transverse momentum and azimuthal angle of the di-lepton pair,
respectively.
For $W^+$ production and decay, we have 

\begin{eqnarray}
 &  & \left({\frac{{\rm d}\sigma(h_{1}h_{2}\rightarrow W^+(\rightarrow\nu_{\ell}\ell^{+}(\gamma))X)}{{\rm d}Q^{2}\,{\rm d}y\,{\rm d}Q_{T}^{2}\,{\rm d}\phi_{W}{\rm \, d}\Pi_{2}}}\right)_{{\rm res}}\nonumber \\
 & = & {\frac{1}{48\pi S}}\Biggl\{\frac{1}{(2\pi)^{2}}\int{\rm d}^{2}b\, e^{i{\vec{Q}_{T}}\cdot{\vec{b}}}\times\sum{\widetilde{W}_{j{\bar{k}}}^{(2)}(b_{*},Q,x_{1},x_{2},C_{1},C_{2},C_{3})}\,\widetilde{W}_{j{\bar{k}}}^{{\rm NP}}(b,Q,x_{1},x_{2})\nonumber \\
 &  & ~~~~~~~~~~~~~~~~~~+Y(Q_{T},Q,x_{1},x_{2},{C_{4}})\Biggl\},\label{eq:ResFor}
 \end{eqnarray}
and
\begin{eqnarray}
 &  & \left({\frac{{\rm d}\sigma(h_{1}h_{2}\rightarrow W^+(\rightarrow\nu_{\ell}\ell^{+}\gamma)X)}{{\rm d}Q^{2}\,{\rm d}y\,{\rm d}Q_{T}^{2}\,{\rm d}\phi_{W}{\rm \, d}\Pi_{3}}}\right)_{{\rm res}}\nonumber \\
 & = & {\frac{1}{48\pi S}}\Biggl\{\frac{1}{(2\pi)^{2}}\int{\rm d}^{2}b\, e^{i{\vec{Q}_{T}}\cdot{\vec{b}}}\times \sum{\widetilde{W}_{j{\bar{k}}}^{(3)}(b_{*},Q,x_{1},x_{2},C_{1},C_{2},C_{3})}\,\widetilde{W}_{j{\bar{k}}}^{{\rm NP}}(b,Q,x_{1},x_{2})\Biggl\},\label{eq:ResFor2}
\end{eqnarray}
where ${\rm {\rm d}\Pi_{2}}$ and ${\rm {\rm d}\Pi_{3}}$ represents
the two-body and three-body phase space of the vector boson decay
products, respectively. In the above equations the parton momentum
fractions are defined as $x_{1}={\displaystyle \frac{Q\: e^{y}}{\sqrt{S}}}$
and $x_{2}={\displaystyle \frac{Q\: e^{-y}}{\sqrt{S}}}$, where $\sqrt{S}$
is the center-of-mass energy of the hadrons $h_{1}$ and $h_{2}$.
The renormalization group invariant quantities $\widetilde{W}_{j\bar{k}}^{(2)}(b)$
and $\widetilde{W}_{j\bar{k}}^{(3)}(b)$, which sum to all orders
in $\alpha_{S}$ all the singular terms that behave as $Q_{T}^{-2}\times\left[1\:{\rm or}\;\ln(Q_{T}^{2}/Q^{2})\right]$
for $Q_{T}\rightarrow0$, are
\begin{eqnarray}
 &  & \widetilde{W}_{j\bar{k}}^{(2)}(b,Q,x_{1},x_{2},C_{1},C_{2},C_{3})\nonumber \\
 & = & e^{{\displaystyle -S(b,Q,C_{1},C_{2})}}\left|V_{jk}\right|^{2}\times \Biggl\{\biggl[(C_{ja}\otimes f_{a/h_{1}})(x_{1})(C_{\bar{k}b}\otimes f_{b/h_{2}})(x_{2})+(x_{1}\leftrightarrow x_{2})\biggl]\nonumber \\
 &  & ~~~~~~~~~~~~~~~~~~~~~~~~~~~~~~~~~~~~~~~~~~~~~~~~\times\frac{{\rm d}\hat{\sigma}_{{\rm F}}^{(0+1)}}{{\rm d}\Pi_{2}}(j\bar{k}\rightarrow\nu_{\ell}\ell^{+}(\gamma))\Biggl\},\label{eq:W}\end{eqnarray}
and
\begin{eqnarray}
 &  & \widetilde{W}_{j\bar{k}}^{(3)}(b,Q,x_{1},x_{2},C_{1},C_{2},C_{3})\nonumber \\
 & = & e^{{\displaystyle -S(b,Q,C_{1},C_{2})}}\left|V_{jk}\right|^{2}\times \Biggl\{\biggl[(C_{ja}\otimes f_{a/h_{1}})(x_{1})(C_{\bar{k}b}\otimes f_{b/h_{2}})(x_{2})+(x_{1}\leftrightarrow x_{2})\biggl]\nonumber \\
 &  & ~~~~~~~~~~~~~~~~~~~~~~~~~~~~~~~~~~~~~~~~~~~~~~~~\times\frac{{\rm d}\hat{\sigma}_{{\rm F}}^{(1)}}{{\rm d\Pi_{3}}}(j\bar{k}\rightarrow\nu_{\ell}\ell^{+}\gamma)\Biggl\},\label{eq:W3}\end{eqnarray}
where 
$\hat{\sigma}_{{\rm F}}^{(0)}$
is the Born level parton cross section for 
$j\bar{k}\rightarrow\nu_{\ell}\ell^{+}$, and 
$\hat{\sigma_{{\rm F}}}^{(1)}$
includes the final-state NLO QED corrections.
The notation $\otimes$ denotes the convolution~\cite{Balazs:1997xd}
 \begin{eqnarray}
(C_{ja}\otimes f_{a/h_{1}})(x_{1}) & = & \int_{x_{1}}^{1}\frac{d\xi_{1}}{\xi_{1}}C_{ja}\left(\frac{x_{1}}{\xi_{1}},b,\mu=\frac{C_{3}}{b},C_{1},C_{2}\right)\times f_{a/h_{1}}\left(\xi_{1},\mu=\frac{C_{3}}{b}\right),\label{eq:convolution}\end{eqnarray}
and the $V_{jk}$ coefficients are the Cabibbo-Kobayashi-Maskawa
mixing matrix
elements. In the above expression $j$ represents quark flavors and
$\bar{k}$ stands for antiquark flavors. The indices $a$ and $b$
are meant to sum over quarks and antiquarks or gluons. Summation on
these double indices is implied. As compared to the results shown
in Refs.~\cite{Balazs:1995nz, Balazs:1997xd}, $\widetilde{W}_{j\bar{k}}^{(2)}(b)$
and $\widetilde{W}_{j\bar{k}}^{(3)}(b)$ contain additional 
$\frac{\alpha}{\pi}$
corrections, which come from the final state QED corrections. The
Sudakov exponent $S(b,Q,C_{1},C_{2})$ in Eqs.~(\ref{eq:W}) and (\ref{eq:W3})
is defined as~\cite{Balazs:1997xd}
\begin{eqnarray}
S(b,Q,C_{1}C_{2}) & = & \int_{C_{1}^{2}/b^{2}}^{C_{2}^{2}Q^{2}}\frac{d\bar{\mu}^{2}}{\bar{\mu}^{2}}\Biggl[A(\alpha_{S}(\bar{\mu}))\ln\left(\frac{C_{2}^{2}Q^{2}}{\bar{\mu}^{2}}\right)+B(\alpha_{S}(\bar{\mu}),C_{1},C_{2})\Biggl].\label{eq:sudakov}\end{eqnarray}
The explicit forms of the $A$, $B$, and $C_{ja}$ functions and
the renormalization constants $C_{i}\:(i=1,2,3)$ can be found in
Appendix D of Ref.~\cite{Balazs:1997xd}. In our calculation, we have
included
$A^{(1)}$, $A^{(2)}$, $B^{(1)}$, $B^{(2)}$, $C^{(0)}$and $C^{(1)}$,
with canonical choice of $C_{i}$'s. The $Y$ piece in Eq.~(\ref{eq:ResFor}),
which is the difference of the fixed order perturbative result and
their singular part, can be found in Appendix E of Ref.~\cite{Balazs:1997xd}.

We follow the prescription in Ref.~\cite{Wackeroth:1996hz}, which
decomposes the electroweak $O(\alpha)$ contribution to the resonant
single $W$ production in a general 4-fermion process into gauge invariant
QED-like and weak parts, to extract a gauge invariant QED-like form
factor from the photon contribution. The NLO FQED differential cross
sections are calculated by using phase space slicing 
method~\cite{Baer:1989jg},
which introduces two theoretical cutoff parameters, soft cutoff $\delta_{{\rm S}}$
and collinear cutoff $\delta_{{\rm C}}$, to isolate the soft and
collinear singularities associated with the real photon emission subprocesses
by partitioning phase space into soft, collinear and hard regions
such that
\begin{equation}
\left|\mathcal{M}^{{\rm r}}\right|^{2}=\left|\mathcal{M}^{{\rm r}}\right|_{{\rm soft}}^{2}+\left|\mathcal{M}^{{\rm r}}\right|_{{\rm collinear}}^{2}+\left|\mathcal{M}^{{\rm r}}\right|_{{\rm hard}}^{2}\,.\label{eq:nlomat}\end{equation}
 The soft region is thus defined by requiring that the photon energy
($E_{\gamma}$) in the ($j\bar{k}$) parton center-of-mass frame 
to satisfy $E_{\gamma}<\delta_{{\rm S}}\sqrt{\hat{s}}/2$,
where $\sqrt{\hat{s}}$ is 
the invariant mass of the ($j\bar{k}$) partons.
 Using the dimensional regularization scheme,
we can then evaluate, in n-dimensions, the real photon emission diagrams
under the soft photon approximation, where the photon momentum is
set to be zero in the numerator, and integrate over the soft region.
In the soft and collinear regions the cross section is proportional
to the Born cross section. The soft singularities originating from
the final-state photon radiation cancel against the corresponding
singularities from the final-state virtual corrections and leave a
finite result depending on the soft cutoff parameter $\delta_{{\rm S}}$.
For $E_{\gamma}>\delta_{{\rm S}}\sqrt{\hat{s}}/2$, the real photon
emission diagrams are calculated in four dimensions using the helicity
amplitude method. The collinear singularities associated with photon
radiation from the final-state charged lepton is regulated by the
finite lepton mass. The end result of the calculation consists of
two sets of weighted events corresponding to the $j\bar{k}\rightarrow\nu_{\ell}\ell^{+}(\gamma)$
and $j\bar{k}\rightarrow\nu_{\ell}\ell^{+}\gamma$ contributions which
are included in Eqs.~(\ref{eq:ResFor}) and (\ref{eq:ResFor2}), separately.
Each set depends on the soft cutoff parameter $\delta_{{\rm S}}$. The sum of
the two contributions, however, is independent of $\delta_{{\rm S}}$,
as long as the soft cutoff is small enough to validate the 
soft-gluon approximation.
In our numerical studies, we take $\delta_{{\rm S}}=0.001$ which yields
a stable numerical result in agreement with 
Refs.~\cite{Dittmaier:2001ay, Baur:1998kt}.
Through our calculation, we adopt the CERN 
LEP line-shape prescription of a resonance state and write 
the $W$ boson propagator as\begin{equation}
\frac{1}{(p^{2}-M_{W}^{2})+iM_{W}\Gamma_{W}\,\, p^2/{M_{W}^2}}\,,\label{eq:runningwidth}\end{equation}
where $\Gamma_{W}$ is the width of $W$ boson.

To examine how much the combined contributions from the initial-state
 QCD resummation and the final-state 
QED corrections can affect the precision measurement of $M_{W}$, we perform
Monte Carlo analyses to study a few experimental observables that
are most sensitive to the measurement of $M_W$ at  
the Tevatron (a $p\bar{p}$ collider with $\sqrt{S}$=1.96
TeV). For the numerical evaluation we chose the following set of SM
parameters: $\alpha=1/137.0359895$, $G_{\mu}=1.16637\times10^{-5}{\rm GeV^{-2}}$,
$M_{W}=80.35$ GeV, $\Gamma_{W}=2.0887$ GeV, $M_{Z}=91.1867$ GeV,
$m_{e}=0.51099907$ MeV. 
Thus, the square of the weak gauge coupling is 
$g^2=4 \sqrt{2} M_W^2 G_\mu$.
Because of the limited space, we focus our attention on
the positively charged electron lepton (i.e. positron) 
only, though our analysis procedure also applies
to the $\mu$ lepton. The complete study including both electron and muon
leptons will be shown in our forthcoming paper, in which we also extend
our study to the LHC. 
\begin{figure}
\includegraphics[scale=0.6]{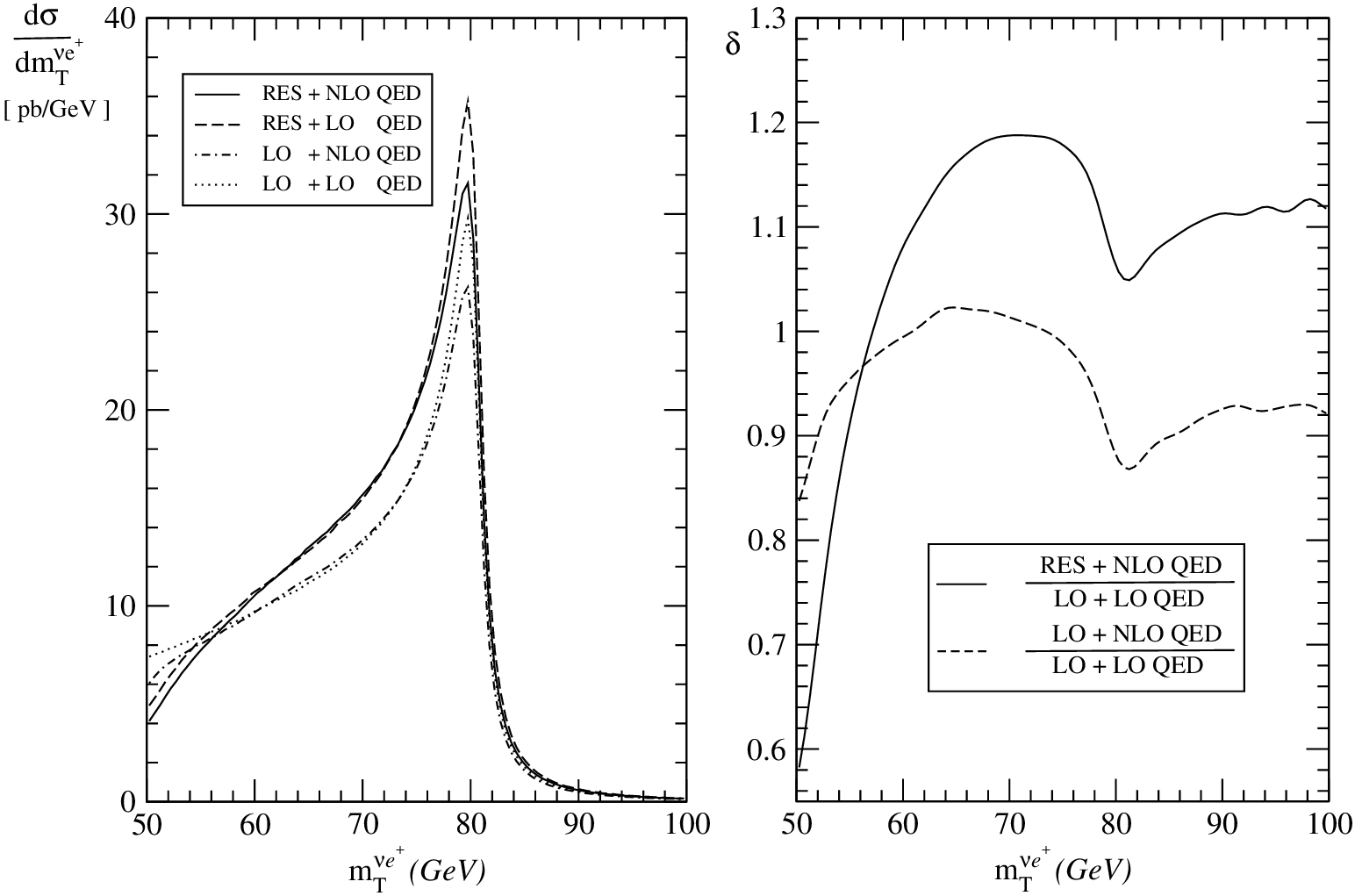}
\caption{Transverse mass distribution of $W^+$ boson \label{fig:tm}}
\end{figure}

The $W$ events in these analyses are selected by demanding a single
isolated high $p_{T}$ charged lepton in conjunction with large missing
transverse energy. To model the acceptance cuts used by the CDF and
\D0 Collaborations in their $W$ mass analyses, we impose the following
transverse momentum ($p_{T}$) and pseudo-rapidity ($\eta$) cuts
on the final-state leptons:
\begin{equation}
p_{T}^{e}>25{\rm GeV,\left|\eta(e)\right|<1.2,\not \! E_{T}>25{\rm GeV}}.\label{eq:basic_cuts}
\end{equation}

Due to the overwhelming QCD backgrounds, the measurement of $W$ boson
mass at hadron collider is performed in the leptonic decay channels. Since
the longitudinal momentum of the neutrinos produced in the leptonic
$W$ boson decays ($W^+\rightarrow\nu_e e^+$) cannot be measured,
there is insufficient information to reconstruct the invariant mass
of the $W$ boson. Instead, the transverse mass distribution of the
final state lepton pair, which exhibits a Jacobian edge at $M_{T}\sim M_{W}$,
is used to extract out $M_{W}$. Transverse mass ($M_{T}^{W}$) of $W$ 
is defined as\begin{equation}
M_{T}^{W}=\sqrt{2p_{T}^{e}p_{T}^{\nu}(1-\cos\phi)} \, ,
\label{eq:trans-mass}\end{equation}
where $\phi$ is the angle between the charged lepton and the neutrino
in the transverse plane. The neutrino transverse momentum 
($p_{T}^{\nu}$) is identified
with the missing transverse energy ($\not \! {\rm E}_{T}$) in the event.
In Fig.~\ref{fig:tm}, we show various theory predictions on
the $M_{T}^{W}$ distribution.
The legend of the figure is defined as follows:
\begin{itemize}
\item LO : including only the Born level initial-state contribution,
\item RES : including the initial-state multiple soft-gluon corrections
via QCD resummation,
\item LO QED : including only the Born level final-state contribution,
\item NLO QED : including the final-state NLO QED corrections.
\end{itemize}
For example, the solid curve (labelled as RES+NLO QED) in 
Fig.~\ref{fig:tm}(a) is the prediction from our combined calculation
given by Eqs.~(\ref{eq:ResFor}) and (\ref{eq:ResFor2}).

As shown in Fig.~\ref{fig:tm}(a), compared to the lowest order cross
section (dotted curve), the initial state QCD resummation effects (dashed
curve) increase the cross section at the peak of the $M_{T}^{W}$ distribution
by about $20\%$, and the final state NLO QED corrections (dot-dashed
curve) decrease it by about $-12\%$, while the combined 
contributions (solid curve) 
of the QCD resummation and FQED corrections increase it by $7\%$.
In addition to the change in magnitude,
the line-shape of the $M_{T}^{W}$ distribution 
is significantly modified by the effects of QCD resummation
and FQED corrections. 
To illustrate this point, we plot the ratio of the (RES+NLO QED)
differential cross sections to the LO ones 
as the solid curve in Fig.~\ref{fig:tm}(b).
The dashed curve is for the ratio of (LO+NLO QED) to LO.
As shown in the figure,
the QCD resummation effect dominates the shape of 
$M_{T}^{W}$ distribution for 
$65\,{\rm GeV\leq M_{W}\leq95\,{\rm GeV}}$, while
the FQED correction reaches its maximal effect around the Jacobian
peak ($M_{T}^{W}\simeq M_{W}$).
Hence, both corrections must be included to accurately predict 
the distribution of $M_{T}^{W}$ around the Jacobian region to determine
$M_W$. We note that after including the effect due to the finite
resolution of the detector (for identifying an isolated electron or
muon), the size of the FQED correction is largely 
reduced~\cite{Dittmaier:2001ay, Baur:1998kt}.
\begin{figure}
\includegraphics[scale=0.6]{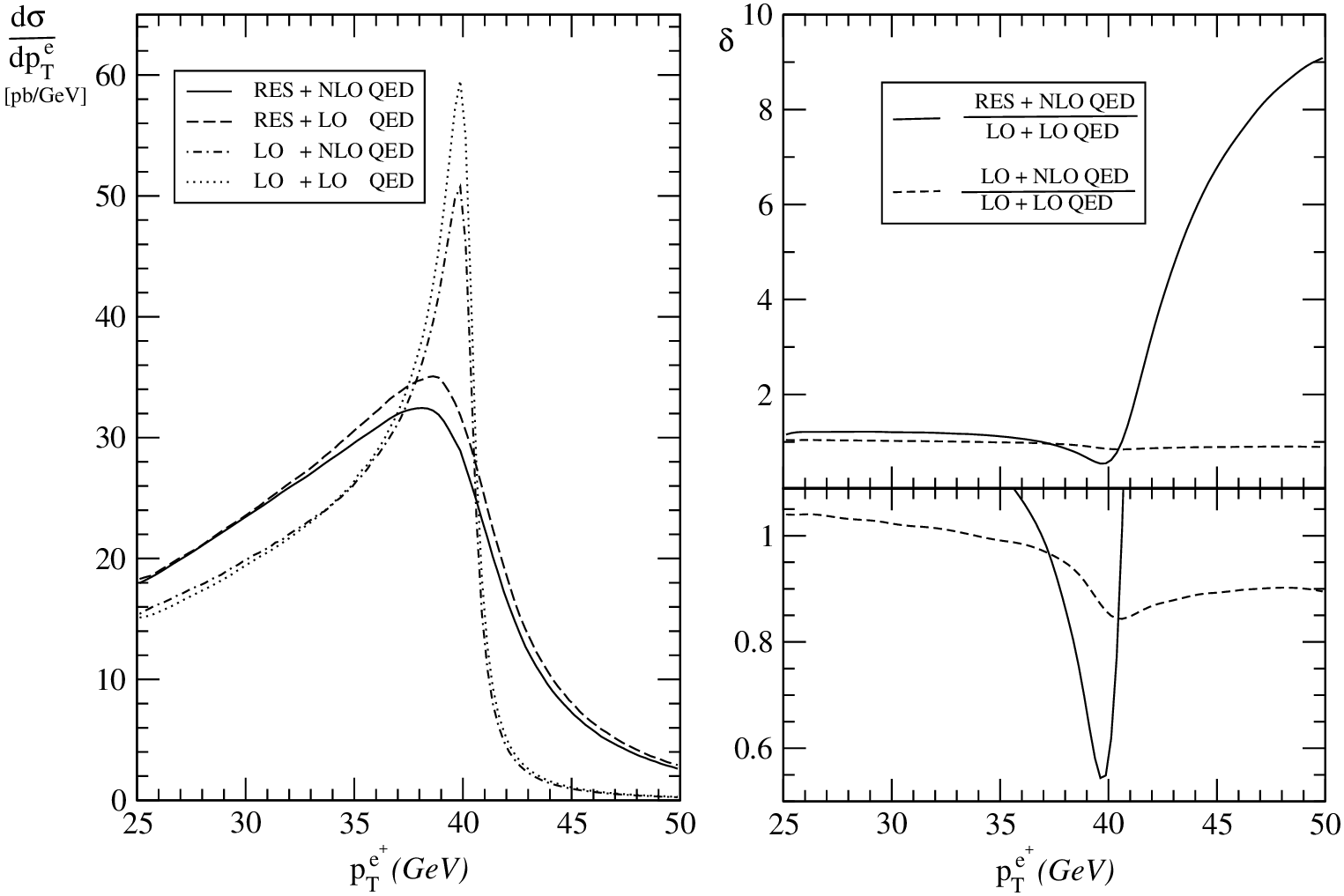}
\caption{Transverse momentum distributions of $e^+$\label{fig:pte}}
\end{figure}

Although the $M_{T}^{W}$ distribution has been the optimal observable 
for determining $M_{W}$ at the Tevatron, it requires 
 an accurate measurement of the missing transverse momentum
direction which is in practice difficult to control. 
On the other hand, the transverse momentum
of the decay charged lepton ($p_{T}^{e}$) is less sensitive to
the detector resolution, so that it can be used to measure $M_W$ 
and provide important cross-check on the result derived from 
the $M_{T}^{W}$ distribution, 
for they have different systematic uncertainties. 
Another important feature of this observable is that
$p_{T}^{e}$ distribution is more sensitive to the transverse momentum 
of $W$ boson. Hence, the QCD soft-gluon resummation effects,
the major source of $p_{T}^{W}$, must be included to reduce
the theoretical uncertainty of this method. 
In Fig.~\ref{fig:pte}(a), we show the $p_{T}^{e}$
distributions predicted by various theory calculations, 
and in Fig.~\ref{fig:pte}(b), the ratios
of the higher order to lowest order cross sections as a function
of $p_{T}^{e}$. The lowest order distribution
(dotted curve) shows a clear and sharp Jacobian peak at 
$p_{T}^{e}\simeq M_{W}/2$,
and the distribution with the 
NLO final-state QED correction (dot-dashed curve) also
exhibits the similar Jacobian peak with the peak magnitude
reduced by about $15\%$. But the clear and sharp Jacobian peak of
the lowest order and NLO FQED distributions 
(in which $p_{T}^{W}=0$)
are strongly smeared by
the finite transverse momentum of the $W$ boson induced by multiple
soft-gluon radiation, as clearly demonstrated by the 
QCD resummation distribution
(dashed curve) and the combined contributions of the QCD resummation and
FQED corrections (solid curve). 
Similar to the $M_{T}^{W}$ distribution, the QCD resummation effect
dominates the whole $p_{T}^{e}$ range, while the FQED correction
reaches it maximum around the Jacobian peak (half of $M_W$).
The combined contribution of the QCD resummation
and FQED corrections reaches the order of $45\%$ near 
the Jacobian peak. Hence, these lead us to conclude that the
QCD resummation effects are crucial in the measurement of $M_{W}$
from fitting the Jacobian kinematical edge of the $p_{T}^{e}$
distribution. 
\begin{figure}[t]
\includegraphics[scale=0.6]{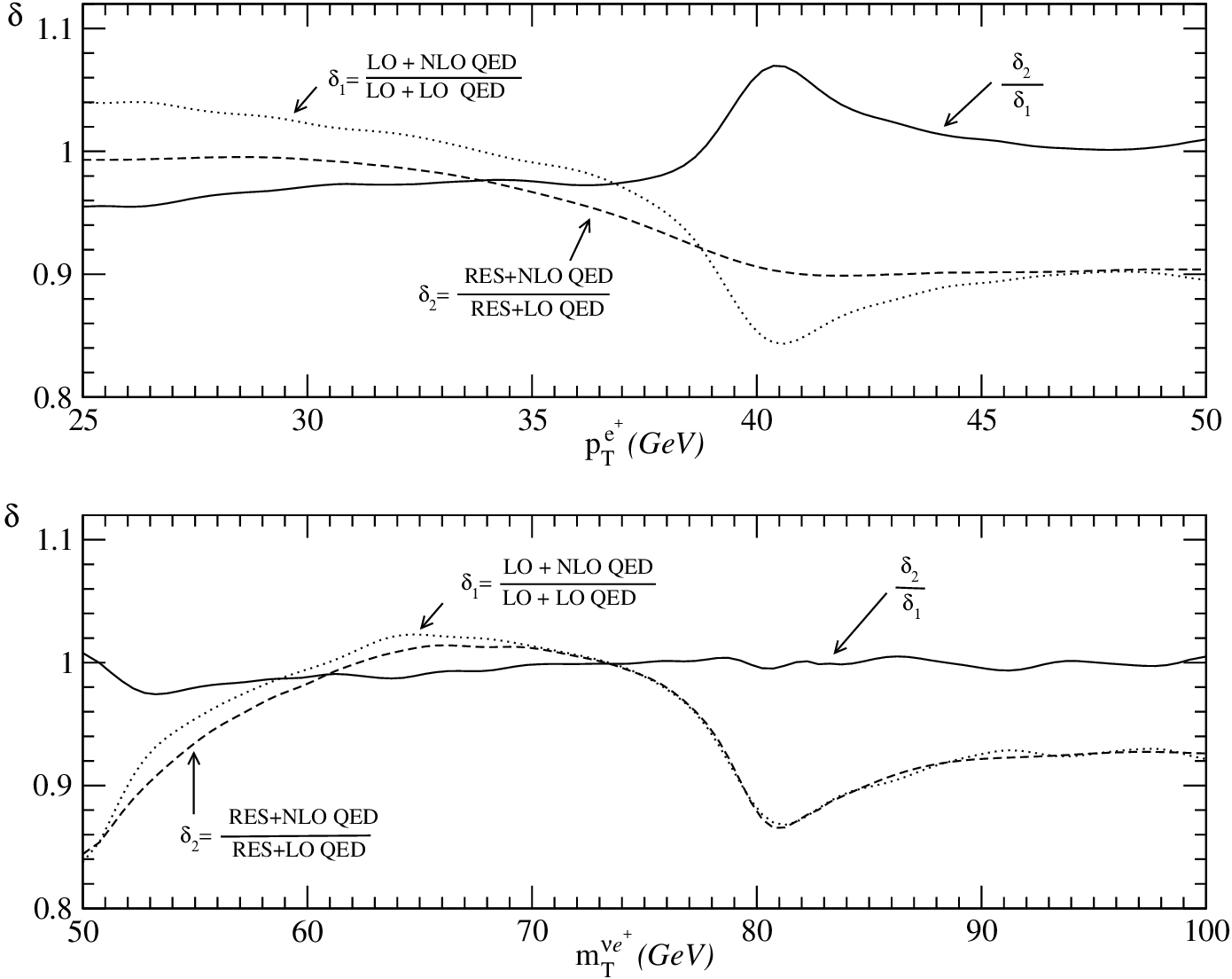}
\caption{Distributions of $\mathcal{R}$, $\delta_1$ and
$\delta_2$\label{fig:ratio}}
\end{figure}

It is also interesting to examine the effect of
the final-state NLO QED correction 
to the theory predictions with LO (leading order) or 
RES (resummed)
initial-state cross sections, 
which is described by the observable $\mathcal{R}$,
defined as
\begin{equation}
{\displaystyle \mathcal{R}=\frac{\delta_{2}}{\delta_{1}}}, \label{eq:R}
\end{equation}
where
\begin{eqnarray}
\delta_{1} & = & \frac{{\rm LO\,+\, NLO\, QED}}{{\rm LO\,+\, LO\,
QED}},\\ \nonumber
\delta_{2} & = & \frac{{\rm RES\,+\, NLO\, QED}}{{\rm RES\,+\, LO\, QED}}. \label{eq:delta1_delta2}
\end{eqnarray}
 The distributions of $\mathcal{R}$ as a function of $p_{T}^{e}$
and $M_{T}^{W}$ are shown 
in the upper part and lower part of Fig.~\ref{fig:ratio},
respectively. As expected, the $\mathcal{R}(M_{T})$ distribution
is almost flat all the way from $60\,{\rm GeV}$ to $100\,{\rm GeV}$,
but the $\mathcal{R}(p_{T}^{e})$ distribution deviates from one
in the Jacobian peak region. This is due to the fact that $p_{T}^{e}$
is more sensitive to $P_{T}^{W}$ than $M_{T}^{W}$.

In order to study the impact of the presented calculation to the determination
of the $W$ boson mass, the effect due to the finite resolution of
the detector should be included, which will be presented elsewhere.

We thank P. Nadolsky and J.W. Qiu for helpful discussions. This work
was supported in part by NSF under grand No. PHY-0244919 and PHY-0100677.


\begin{thebibliography}{99}
\bibitem{RESBOS-A}A Fortran code that implemets the theoretical calculation presented in this work.
\end{thebibliography}
\end{document}